%% file: IFACPaper.tex
\documentclass{ifacconf}

\usepackage{graphicx}      
\usepackage{natbib}        
\usepackage{amsmath}
\usepackage{bm}
\usepackage{amssymb}
\usepackage{siunitx}
\usepackage{subcaption}

\input{math_commands}

\input{pace_shorthands}

\begin{document}
\begin{frontmatter}

\title{Horizon Selection in Physics-Enhanced Neural ODEs: Theoretical Insights and Flux Linkage Application}


\author[First]{Giulio Montecchio}
\author[First]{Benjamin Hartmann}
\author[First]{Sven Reimann}
\author[First]{Maximilian Manderla}
\author[First]{Jan Achterhold}
\author[Second]{Daniel Görges}

\address[First]{Robert Bosch GmbH, Renningen, Germany (e-mail: first.last@de.bosch.com)}
\address[Second]{RPTU University Kaiserslautern-Landau, Germany}

\begin{abstract} 
The integration horizon during the training plays a critical role in Physics-Enhanced Neural Ordinary Differential Equations. We draw conclusions about horizon extension in the training of Neural Ordinary Differential Equations based on classical nonlinear system identification of input-output models. In light of this insight, we propose a framework that exploits longer horizons to reduce bias in physical parameter estimates, extracts residual information from data, and acts as a regularizer improving generalization. In the learning of a model for permanent magnet synchronous machine, the method is used to jointly estimate the flux map and the resistance.
\end{abstract}


\end{frontmatter}

\section{Introduction}
Neural Ordinary Differential Equations (Neural ODEs) have experienced a recent resurgence thanks to advances in machine learning engineering and theory, exemplified by works such as \citep{Chen_Rubanova_Bettencourt_Duvenaud_2019}.~\cite{Rackauckas_Ma_Martensen_Warner_Zubov_Supekar_Skinner_Ramadhan_Edelman_2021} introduced the broad concept of Universal Differential Equation, which provides a versatile tool for hybrid modeling by coupling mechanistic differential equations with data-driven components.
Following~\cite{Sorourifar}, we adopt the term Physics-Enhanced Neural ODEs (PeN-ODEs) to refer specifically to ordinary differential equations that integrate Neural Network (NN) components with explicit first-principle dynamical relations. This terminology highlights the explicit embedding of known physics into the model, combining NN representational power with physical constraints to improve extrapolation, data efficiency, interpretability, and compatibility with existing control structures.

In this paper, we investigate the crucial role of the integration horizon during PeN-ODE training. While being treated as a hyperparameter, its selection profoundly influences the training dynamics. Our novel contribution is to delineate an explicit connection between the training of Neural ODE and established concepts in the field of nonlinear system identification, specifically Nonlinear Autoregressive with Exogenous Input (NARX) and Nonlinear Output Error (NOE) predictors. Based on this connection, we elucidate the role of the integration horizon and develop a framework to mitigate estimation bias of the physical parameters in a PeN-ODE. To the best of the authors' knowledge, this explicit theoretical nexus and its practical implications have not been thoroughly discussed in this context before.

The practical relevance of the integration horizon is then demonstrated through a real-world application: the identification of the flux linkage map and resistance of a Permanent Magnet Synchronous Machine (PMSM) from experimental measurement data. While PMSM parameter identification via PeN-ODE was already proposed in \citep{Sevcik_Smidl_Glac_2025}, our work specifically leverages this application to illustrate how an increased integration horizon can mitigate bias arising from inherent model-plant mismatch.

The theoretical insights into the connection between Neural ODE training and NARX models are explained in Section 2. In Section 3 we exploit this connection to provide a principled approach for the choice of the training horizon. Section 4 presents the PeN-ODE framework for PMSM parameter identification, and the experimental results are discussed in Section 5. In Section 6 we draw conclusions and describe outlooks.

\section{Integration and prediction}
Let us denote a general ODE as
\begin{equation}
\label{eq:ODE}
    \dot{\bm x} = f( \bm x, \bm u, \bm \theta)
\end{equation}
where $\bm x \in \mathbb{R}^n$ is the measurable state vector, $\bm u \in \mathbb{R}^m$ is the input vector, and $\bm \theta \in \mathbb{R}^p$ are the parameters of the model we want to identify, for example the weights of an NN. The function $f$ describes the dynamics of the system. It can combine an NN with relations coming from first principles, forming what has been defined as a PeN-ODE. \par 
The training of such a structure is invariably linked to the choice of an ODE solver, which allows the model to be trained based on an existing dataset
\begin{equation}
\label{eq:dataset}
\mathcal{D} = \{ (\bm u_k, \bm x_k) \}_{k=0}^N.
\end{equation}
Here, $\bm u_k$ represents the input applied at time $t_k$, and $\bm x_k$ is the corresponding measured state at time $t_k$.
To better describe neural ODEs with classic system identification terminology, we can rewrite the integration step of an ordinary differential equation as a discrete-time predictor, connecting to the pivotal model representation of~\cite{Ljung_1999}. This perspective is particularly useful as it allows us to benefit from the insights of system identification techniques for model training. With this purpose, the operations performed by an explicit single-step ODE solver with a fixed step size can be written as a nonlinear predictor
\begin{equation}
\label{eq:explicit_singlestep_integrator_predictor}
    \hat{\bm x}_{k+1} = g(\bm x_k,\bm u_{k}, \bm \theta),
\end{equation}
where $\hat{\bm x}_{k+1}$ is the state obtained after the integration step. This writing collects the dynamics $f$ and the operations of the ODE solver (like intermediate evaluations of $f$ within the time step) into a single function~$g$. Equation~\eqref{eq:explicit_singlestep_integrator_predictor} can then be thought of as a NARX, as defined, for example, in~\cite[Chapter 19]{Nelles_2020}.
\subsubsection{Example.} Solving the ODE~\eqref{eq:ODE} with the Forward Euler method and a fixed step size $T$ equal to the sampling interval, the predictor~\eqref{eq:explicit_singlestep_integrator_predictor} takes the form
$$ g(\bm x_k, \bm u_k, \bm \theta) = \bm x_k + T \cdot f(\bm x_k,\bm u_k, \bm \theta). $$
\subsection{Generalization and limitations of the NARX model}
Here, we discuss the generality of the NARX model with respect to the solver choice. Further details on the solver classification can be found in \citep{Butcher_2008}. \par
The most commonly used fixed-step methods for Neural ODEs, such as Forward Euler or explicit methods from the Runge-Kutta family, result in~\eqref{eq:explicit_singlestep_integrator_predictor}.
Generalization to multistep integrators, such as Adams-Bashforth methods, is straightforward. These methods use multiple past values of $\bm x_k$ and $\bm u_k$ to perform one solver step. These past values can be directly fitted into the NARX structure
\begin{equation}
\label{eq:explicit_multistep_integrator_predictor}
    \hat{\bm x}_{k+1} = g(\bm x_k, \ldots, \bm x_{k-n_x}, \bm u_{k}, \ldots, \bm u_{k-n_u}, \bm \theta).
\end{equation}
The generalization to variable step size integration methods would be possible by considering time-varying predictors, but is out of the scope of this paper. While explicit solvers lead to an explicit predictor, implicit ODE solvers would result in an implicit equation. An implicit single-step solver could be written as
\begin{equation}
\label{eq:implicit_singlestep_integrator_predictor}
    \hat{\bm x}_{k+1} = g(\hat{\bm x}_{k+1}, \bm x_k, \bm u_{k+1}, \bm u_{k}, \bm \theta)
\end{equation}
and is no longer a NARX. However, such a structure would make both the training and the application of the learned predictor more challenging, and is therefore left out of the investigation. 
\subsection{Integration as $k$-step-ahead prediction}
The NARX predictor described by either~\eqref{eq:explicit_singlestep_integrator_predictor} or~\eqref{eq:explicit_multistep_integrator_predictor} represents a single integration step. Neural ODE training can exploit time-series data by integrating the model over longer horizons, i.e., for more than one step. In this scenario, the integration can be naturally reformulated as a recurrent structure. \\
If we perform an integration such as~\eqref{eq:explicit_singlestep_integrator_predictor} over a horizon of $H$ time steps from the dataset $\mathcal{D}$ and we start from the initial condition $\bm x_0$, the subsequent integrated states $\hat{\bm x}_k$ for $k=1, \ldots, H$ are generated recurrently from 
\begin{equation}
\label{eq:multistep_prediction}
    \hat{\bm x}_{k+1} = g( \hat{\bm x}_{k}, \bm u_{k}, \bm \theta) \quad \text{for } k= 0, \ldots, H-1
\end{equation}
where $\hat{\bm x}_0 = \bm x_0$. Such a structure is described as $k$-step-ahead predictor in \citep{Ljung_1999}. 
 The formula in~\eqref{eq:multistep_prediction} can be straightforwardly adapted to represent the 
 integration with a multistep integrator by recurrently feeding the predictor~\eqref{eq:explicit_multistep_integrator_predictor}, provided with enough past states and inputs for the initial conditions. \par
We now introduce the NOE predictor:
\begin{equation}
        \hat{\bm x}_{k+1} = g( \hat{\bm x}_k, \ldots, \hat{\bm x}_{k-n_x}, \bm u_{k}, \ldots, \bm u_{k-n_u}, \bm \theta).
\end{equation}
As stated in~\cite[p. 162]{Ljung_1999}, the only difference between NOE and $k$-step-ahead predictors is that the latter makes use of the measured initial conditions.
The effects of optimizing a $k$-step-ahead predictor have been widely studied, under various names, including output error method, simulation error method, parallel training, MPC relevant identification. In particular, \citep{Farina_Piroddi_2008} examines how, as the horizon increases, the properties of the model obtained with $k$-step-ahead prediction minimization tend to align with the output error method, and \citep{Aguirre_Barbosa_Braga_2010} offers a practical study of the convergence and bias properties on abstract model examples. 
\subsection{Training of the Neural ODEs}
The training of the Neural ODE usually proceeds by minimizing the squared error on the predicted states; see \citep{Kidger_2022}. This optimization is a supervised learning process, where the measured state $\bm x_k$ serves as the label. Given a chosen prediction (or integration) horizon $H$ and a set of $N_{\mathrm{IC}}$ initial conditions (which are measured states from the dataset $\mathcal{D}$), the training loss can be written as
\begin{equation}
\label{eq:mse_loss_function}
    L(\bm \theta) = \frac{1}{HN_{\mathrm{IC}}}\sum_{j=1}^{N_{\mathrm{IC}}} \sum_{k=0}^{H-1} \| \bm x_{k+1}^{(j)} - g( \hat{\bm x}_{k}^{(j)}, \bm u_{k}, \bm \theta) \|^2,
\end{equation}
where $\bm x_{k}^{(j)}$ is the measured state at time $t_k$ corresponding to the $j$-th initial condition, and $\hat{\bm x}_{k}^{(j)}$ is the predicted state generated by the integration~\eqref{eq:multistep_prediction}, which is for each sequence $j$ initialized as $ \hat{\bm x}_0^{(j)} = \bm x_0^{(j)}$. The loss corresponds to the mean squared error of the $k$-step-ahead predictor over the horizon $H$ and the $N_{\mathrm{IC}}$ initial conditions. In the light of the connection shown here, the training of a Neural ODE can be interpreted as the identification of a one-step-ahead predictor, i.e., the NARX in~\eqref{eq:explicit_singlestep_integrator_predictor} or~\eqref{eq:explicit_multistep_integrator_predictor}. The integration horizon $H$ controls the trade-off between properties associated with NARX training ($H=1$) or with NOE training ($H>1$). For an extensive explanation, the reader may refer to~\cite[Chapter 19]{Nelles_2020}.

\section{Horizon Selection}
In the following, we explore two key rationales guiding the selection of the integration horizon, that directly come from nonlinear input-output model optimization: the purpose of the model and the minimization of the estimation bias. 
\subsection{Purpose of the model}
The primary consideration when choosing the optimization criterion is the intended use of the learned model. We can broadly categorize model usage into two scenarios, as in~\cite[Chapter 19]{Nelles_2020}:
\begin{enumerate}
    \item \textbf{Simulation or Open-Loop Operation:} Using the model for long-term simulation, such as in a Model Predictive Control or as an open-loop virtual sensor. In this case, the model operates recurrently, feeding its own predictions back as inputs for subsequent steps.
    \item \textbf{Closed-Loop Components:} Embedding the learned model within blocks of a control architecture that are continuously fed with the latest available measurements, such as in observers or classic feedback controllers. Here, the model primarily performs one-step-ahead predictions.
\end{enumerate}
In classical control theory, these two usages are often associated with two different training approaches: training like a NOE model for the first scenario, or training like a NARX model for the second. 
Training a model like a NOE addresses potential error accumulation in simulation, but the recurrent output leads to a highly non-convex optimization problem.
 Conversely, training a model like a NARX provides the optimal one-step-ahead prediction. If the predictor is linear in the parameters, this optimization problem is convex, which is why it is so widely adopted.

\subsection{Bias minimization}
For one-step-ahead prediction (i.e., $H=1$), training this predictor leads to an asymptotically optimal and unbiased estimate of the true system dynamics under certain conditions. These include:
\begin{enumerate}
    \item The noise contaminating the measured output is white, Gaussian, and independent of past inputs and states.
    \item The characteristics of the noise contaminating the measured output are accurately identified with a proper noise model, which is equivalent to an appropriate pre-filtering of the data, as stated in \citep{Ljung_1999}.
\end{enumerate}
However, this unbiasedness breaks down when there is a significant mismatch between the model's structure and the true plant dynamics, which is often the case for simplified models in control architectures or when incomplete physical knowledge is available.
Based on the affinity between NARX training and PeN-ODEs training established in Section 2, we argue that an increased integration horizon can serve an additional, crucial purpose in PeN-ODE training, especially when the model's structure~\eqref{eq:ODE} cannot fully represent the underlying process. Under such model-plant mismatch, optimizing solely for one-step-ahead prediction ($H=1$) can lead to a biased and inconsistent parameter estimate. Larger values of $H$ introduce a trade-off: they result in a more complicated, non-convex optimization problem, but they offer the possibility of mitigating this bias. By forcing the model to predict accurately over a longer range, even with structural deficiencies, the optimization can learn parameters that are more robust to the model's inaccuracies. This potential to overcome bias by training with longer horizons is a significant advantage in PeN-ODEs, where learnable black-box structures are combined with physical parameters to achieve the most accurate and physically consistent representations possible.

\section{PeN-ODE for PMSM Identification} \label{s: Application}
Permanent Magnet Synchronous Motors are widely used, and accurate knowledge of their electrical parameters, especially the magnetic flux linkages, is crucial for high-end control applications, including optimal current trajectory generation (maximum torque per Ampere, flux weakening), robust state observers, model predictive control, parameter tuning, and high-fidelity drive simulation. This section details how the PeN-ODE framework identifies the flux map and resistance of a PMSM. This framework was introduced in \citep{Sevcik_2023} and extended in \citep{Sevcik_Smidl_Glac_2025}, with a focus on improving model accuracy. With a deeper understanding of the impact of the horizon, we can now exploit the same framework to overcome fundamental limitations of existing estimation methods.

\subsection{PMSM dq-model and flux linkages}
The dynamic behavior of a PMSM is best described in the synchronously rotating dq-reference frame. This reference frame simplifies control design and is the basis for high-performance control architectures such as Field-Oriented Control and for robust torque observers. The voltage equations in this frame are
\begin{equation}
\label{eq:continuous_dq_equation}
    \udq = R \idq + \omega J \psidq + \ddt \psidq,
\end{equation}
where $R$ is the stator resistance (assumed symmetric on the three phases), $\omega$ is the electrical angular velocity of the rotor, $J = \begin{bmatrix} 0 & -1 \\ 1 & 0 \end{bmatrix}$ is the skew-symmetric rotation matrix, and $\psidq = [\psi_\mathrm{d}, \psi_\mathrm{q}]^T$ is the stator flux linkage vector, with $\ddt \psidq$ representing its time derivative.

In this frame, the three-phase stator voltages $\uabc$ and currents $\iabc$ are transformed into their dq-components. Specifically, they are derived from the three-phase stator voltages and the electrical rotor angle $\varphi$ using the amplitude-invariant Clarke-Park transform
\begin{equation}
\nonumber
    \udq = \frac{2}{3} \begin{bmatrix} \cos(\varphi) & \cos(\varphi - \frac{2\pi}{3}) & \cos(\varphi + \frac{2\pi}{3}) \\ -\sin(\varphi) & -\sin(\varphi - \frac{2\pi}{3}) & -\sin(\varphi + \frac{2\pi}{3}) \end{bmatrix} \uabc.
\end{equation}
 To be more precise, $\uabc$ are the actual control inputs, and $\varphi$ can be treated as a known disturbance. Hereafter, for the sake of brevity, we refer to only the dq-quantities, with the understanding that they are transformed from the three-phase system using the measured rotor angle $\varphi$. \par
The flux linkages, central to the machine's magnetic modeling, are typically represented as a parametrized nonlinear function of the currents:
\begin{equation}
\label{eq:psidq_as_f_of_variables}
    \psidq = \boldsymbol{\psi}(\idq, \bm{\theta}).
\end{equation}
The structure of $\boldsymbol{\psi}$ is chosen to accurately capture complex phenomena such as magnetic saturation and cross-saturation, which are critical for precise control. While extending this function to further inputs (such as speed $\omega$ or the electrical angle $\varphi$) is possible and even advised for higher fidelity, it is omitted here for simplicity.
\subsection{Conventional flux linkage identification}
\label{s: conventional_methods}
As flux linkages are typically not directly measurable, they are commonly estimated from measured voltages $\udq$, currents $\idq$, rotor angle $\varphi$, and velocity $\omega$. A distinction can be made between methods based on either stationary or transient experiments. In the former, data is averaged at stationary operating points, while in the latter, the measured quantities are utilized within a discretized form of~\eqref{eq:continuous_dq_equation}.
Many methods based on transient experiments implicitly or explicitly reformulate the problem into minimizing a one-step-ahead prediction error. For instance,~\cite{Wiedemann_Hackl_2023} explicitly shape their model as a one-step-ahead current prediction minimization, while the methods presented in \citep{Montecchio_2025}  and \citep{Ortombina_Pasqualotto_Tinazzi_Zigliotto_2020} implicitly involve the optimization of a one-step-ahead flux predictor such as 
\begin{equation}
\label{eq:flux_predictor}
\hat{\bm{\psi}}_{\mathrm{dq},k+1} = g \left( \bm{i}_{\mathrm{dq},k}, \Bar{\bm u}_{k}, \bm \theta \right),
\end{equation}
where $\Bar{\bm u}_{k} = \left[ \bm{u}_{\mathrm{dq},k},  \omega_{k} \right]^T$ combines the known inputs and disturbances. For linear models and linear one-step-solvers, this (second) type of problem typically leads to a convex optimization.
Such a formulation is inherently affected by the limits of NARX training, and cannot be directly simulated recurrently, as its output is a flux, and not a current. 

\subsection{PeN-ODE formulation}
To leverage the advantages of Neural ODEs, we reformulate the voltage equation~\eqref{eq:continuous_dq_equation} into a state-space ODE for the currents $\idq$. To achieve this, we use the chain rule for the flux derivative $\frac{d}{dt} \psidq = \bm{L}(\idq, \bm \theta) \frac{d}{dt} \idq$, where $\bm{L}(\idq, \bm \theta)$ is the Jacobian of the flux map with respect to the state $\idq$, also known as the differential inductance matrix, and is written as
\begin{equation}
\bm L(\idq, \bm \theta) = \frac{\partial \boldsymbol{\psi}}{\partial \idq} = \begin{bmatrix} \frac{\partial{\psi_\mathrm{d}(\idq, \bm \theta)}}{\partial{\id} } & \frac{\partial{\psi_\mathrm{d}(\idq, \bm \theta)}}{\partial{\iq} } \\ \frac{\partial{\psi_\mathrm{q}(\idq, \bm \theta)}}{\partial{\id} } & \frac{\partial{\psi_\mathrm{q}(\idq, \bm \theta)}}{\partial{\iq} } \end{bmatrix}.
\end{equation}

Substituting this into~\eqref{eq:continuous_dq_equation} and rearranging for $\ddt \idq$ yields 
\begin{equation}
    \label{eq:continuous_dq_current_ODE}
    \ddt \idq = \bm{L}^{-1}(\idq, \bm \theta) \left[ - R \idq - \omega \boldsymbol{J} \boldsymbol{\psi}(\idq, \bm \theta) + \udq \right].
\end{equation}

This formulation fits the general ODE form $\dot{\bm x} = f(\bm x, \Bar{\bm u}, \Bar{\bm \theta})$, with $\bm x = \idq$, $\Bar{\bm u} = \left[\udq, \omega \right]^T$, and $\Bar{\bm \theta} = \left[R, \bm \theta \right]^T$. In this PeN-ODE we have a combination of a black-box model for the flux map, with uninterpretable parameters, and a physical parameter. Extensions of this ODE are common to represent further effects on the electric drive, as is done by~\cite{Wiedemann_Hackl_2023} and~\cite{Montecchio_2025}.
With the choice of an ODE solver, we can write the integration of~\eqref{eq:continuous_dq_current_ODE} as
\begin{equation}
\label{eq:current_predictor}
\hat{\bm{i}}_{\mathrm{dq},k+1} = g \left( \bm{i}_{\mathrm{dq},k}, \Bar{\bm u}_{k}, \Bar{\bm \theta} \right).
\end{equation}
The PeN-ODE approach directly learns the nonlinear flux map $\boldsymbol{\psi}(\idq, \bm \theta)$ within the fundamental predictor structure, and is therefore optimized for the specific ODE solver.

\section{Experimental results}
In this section, we show how a longer horizon during training can reduce the bias of the estimation and improve the model's generalization (i.e., how well it predicts current on different datasets and in various scenarios). \par 
The data
 \begin{equation}
    \mathcal{D} = \{ (\Bar{\bm u}_k, \bm x_k) \}_{k=0}^N =  \{ \left(\left[ \bm{u}_{\mathrm{dq},k}, \omega_k \right], \bm{i}_{\mathrm{dq},k} \right) \}_{k=0}^N
\end{equation}   
are collected from two different systems: 
\begin{itemize}
    \item A simulation model based on a finely gridded multi-dimensional look-up table of the inductance. The model also entails higher harmonics of the flux linkages, which are both current- and angle-dependent. Such behavior cannot be fully represented by an $\idq$-dependent flux map.
    \item A PMSM mounted on a test-bench directly coupled with a load machine that controls the speed. Data are affected by disturbances and noise from different sources and of various natures.
\end{itemize}
In both systems, the model we chose cannot fully represent the plant. For this reason, the joint estimation based on classic one-step-ahead prediction minimization is biased, and the estimation may converge to the wrong physical parameters if the experiment is not designed carefully. We demonstrate that increasing the horizon reduces this bias, using two transient experiments where the method presented in \citep{Montecchio_2025} for joint identification of flux map and resistance previously failed. \par
\subsection{Model and experiment design}
We train a PeN-ODE to learn the resistance and a neural network for the flux linkage. The neural network is designed with the architecture from~\cite{Sevcik_Smidl_Glac_2025}: a fully connected feedforward network that maps $\idq$ currents to $\psidq$ fluxes, featuring two hidden layers of 6 and 4 neurons and employing 
hyperbolic tangent activation functions. \par
The experiment consists of a transient trajectory in the dq current disc, limited by the maximum nominal current $I_\text{max}$ and at constant motor speed. The trajectory is tracked by an approximately tuned current controller. The reference currents are generated using a Lissajous trajectory (also known as a double-sine sweep), and computed as
\begin{align}
i_{\mathrm{q},\mathrm{ref}} &= I_{\mathrm{max}} \sin\left({\phi_\mathrm{q}} t\right) \\
i_{\mathrm{d},\mathrm{ref}} &= \sqrt{I_{\mathrm{max}}^2 - i_{\mathrm{q},\mathrm{ref}}^2} \sin\left(\phi_\mathrm{d} t\right),
\end{align}
where the angular frequencies $\phi_\mathrm{d}$ and $\phi_\mathrm{q}$ are design parameters of the trajectory.
According to the machine's use case, the identification could be limited to subdomains of the disc of maximum current. For example, in motor operation, the identification could be limited to the positive $i_\mathrm{q}$ and negative $i_\mathrm{d}$ domains. \par
The experiments are then conducted with the following parameters:
\begin{itemize}
    \item Simulation: 1500 rpm, $\phi_\mathrm{d}=14 $ rad/s, $\phi_\mathrm{q}=15$ rad/s, 2 s, sampled at 10 kHz.
    \item Test-bench: 1000 rpm, $\phi_\mathrm{d}=\frac{49}{12}$ rad/s, $\phi_\mathrm{q}=\frac{50}{12}$ rad/s, 6 s, sampled at 10 kHz.
\end{itemize}

\subsection{Training strategy}
The training algorithm, implemented in PyTorch, leverages \citep{torchdiffeq} for efficient integration and Adam optimizer from~\cite{kingma2014adam} with default values to better deal with the highly non-convex optimization. For each horizon $H$, the ordered dataset $\mathcal{D}$ is split into sequences of length $H+1$. We employ a Runge-Kutta 4 ODE solver with a fixed step size of $T = 0.1$ ms, matching the sampling time of the experiments. To prevent divergence of suboptimal models along the integration horizon, especially for longer horizons, training is warm-started by iteratively increasing the horizon over the initial epochs. Resistance $R$ is randomly initialized between 0 and 20 m$\Omega$. Convergence is accelerated by assigning a higher learning rate to $R$ compared to $\bm \theta$.

\begin{figure}[htbp] 
    \centering
    \begin{subfigure}[b]{0.48\textwidth} 
        \centering
        \includegraphics[width=\textwidth]{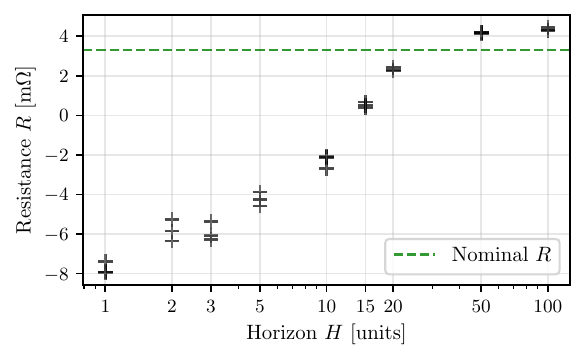}
        \caption{Estimated resistance $R$.}
        \label{fig:R_S2}
    \end{subfigure}
    \hfill 
    \begin{subfigure}[b]{0.48\textwidth}
        \centering
        \includegraphics[width=\textwidth]{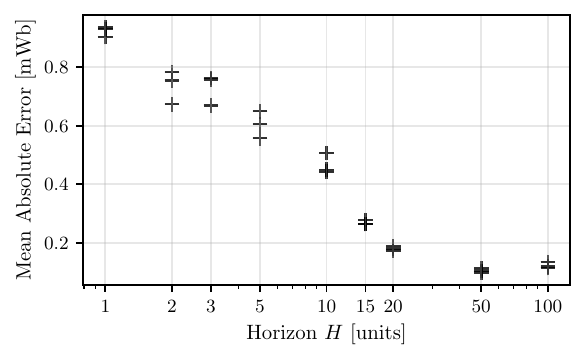}
        \caption{MAE of the estimated fluxes $\psidq$.}
        \label{fig:flux_MAE_S2}
    \end{subfigure}
    \caption{Accuracy of the resistance (a) and flux linkage (b) estimated from simulation.}
    \label{fig:combined_S2_results} 
\end{figure}

\begin{figure}
\begin{center}
\includegraphics[width=8.4cm]{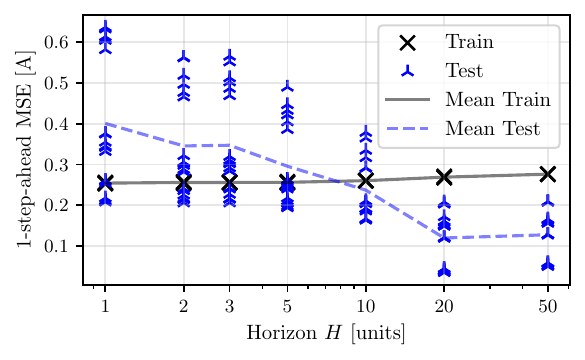}
\caption{ Mean squared error (MSE) of the one-step-ahead $\idq$ prediction from simulation. Each marker denotes the error computed on a different dataset. All datasets are collected at different operating conditions.} 
\label{fig:S2_bias_tradeoff}
\end{center}
\end{figure}

\begin{figure}
    \centering
\includegraphics[width=8.4cm]{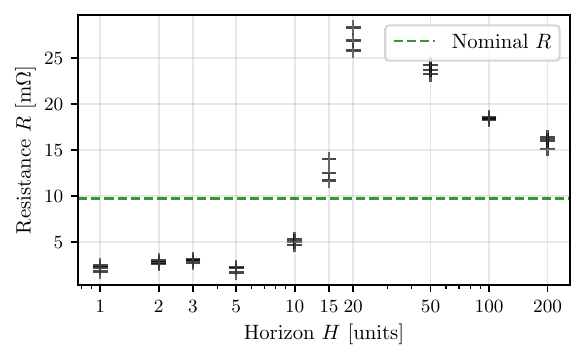}
\caption{Estimated resistance $R$ from test-bench.} 
\label{fig:R_TB}
\end{figure}

\subsection{Results}

Figure \ref{fig:R_S2} presents the estimated resistance $R$ for various horizons $H$, with each marker representing a distinct training run (different random initialization and batch sequencing). We observe consistent convergence of $R$ estimates to the same region for a given $H$ across three different training realizations. Notably, for horizons $H<15$, the estimated $R$ is negative, indicating a non-physical result. This error can be attributed to bias: while this resistance minimizes the one-step-ahead prediction error, its estimate is biased due to the model-plant mismatch. Crucially, increasing the horizon effectively reduces the estimation error of $R$.
In Figure \ref{fig:flux_MAE_S2} we can see that the flux estimate benefits from the increased horizon as well. The mean error, computed over both $\psid$ and $\psiq$ values, decreases. This quantity is available only in simulation, where a ground truth of the flux can be inferred from the look-up tables.

Further insight into the impact of $H$ is provided by Figure \ref{fig:S2_bias_tradeoff}. This figure illustrates the generalization performance of models trained at each horizon across diverse test trajectories (varying scenarios, speeds, and currents). It clearly demonstrates that larger horizons lead to better model generalization to unseen datasets, at the price of slightly worse performance on the training data. Thus, the choice of horizon can be interpreted as a regularization technique, effectively mitigating overfitting.

When considering real test-bench data, assessing the benefits of an increased horizon by observing the resistance estimate is not straightforward. As depicted in Figure \ref{fig:R_TB}, achieving an unbiased resistance estimate is challenging. For certain horizon choices, an increase (e.g., from $H=15$ to $H=20$) can paradoxically worsen the resistance estimates, causing the predictor to perform worse than those found with shorter horizons. Nevertheless, with further increases in the horizon, the resistance estimate appears to improve again. This complex behavior likely stems from the higher number of unmodeled dynamics and signal imperfections affecting the real machine, as no signal conditioning or correction was applied before identification. This decision was made to test the method's recovery capability under the harshest practical conditions. The system is known to be affected by unbalanced currents, sensor errors, current-dependent harmonics in the resolver, significant speed noise, and inverter distortions. A comprehensive understanding of why these specific model mismatches disproportionately affect certain horizons, along with strategies to address them, is reserved for future research.

The method was also evaluated on properly designed experiments (i.e., experiments at low speed and highly dynamic), in which an accurate joint estimation is feasible already using one-step-ahead prediction error minimization. These tests were conducted both in simulation and on a test-bench. We observed no deterioration in the estimated values. However, the larger horizon increases optimization complexity, so the extra computational cost is not justified by a performance gain. This finding agrees with the conclusions of~\cite{Sevcik_Smidl_Glac_2025}, which reports no significant effect from increasing the horizon beyond $H=2$. Thus, horizon extension should be viewed primarily as a means to extract any remaining information from imperfect data rather than as a way to improve estimation accuracy.

\section{Conclusion}
We have shown that the choice of integration horizon during PeN-ODE training has fundamental implications for estimator bias and generalization. By framing explicit-step ODE solvers as NARX predictors and the recurrent integration as a $k$-step-ahead (NOE-like) predictor, the integration horizon is revealed as the key knob that trades off one-step optimality with long-term simulation fidelity. We suggest that longer horizons should be used not only when the target of the model is simulation, but also when the goal is to estimate physically accurate parameters. Training with longer horizons reduces bias in the estimation of physical parameters (demonstrated on PMSM flux map and resistance estimation) and improves out-of-distribution prediction performance, while making the optimization more non-convex and less robust. Practically, the problem can be improved by employing warm-start horizon schedules, and assigning tailored optimization settings to physical parameters. Future work will aim to develop automated horizon-selection and regularization methods to balance bias and training tractability.
\section*{DECLARATION OF GENERATIVE AI AND AI-ASSISTED TECHNOLOGIES IN THE WRITING PROCESS}
During the preparation of this work the authors used an internal server of LLMs in order to rephrase, find synonyms, and proofread. After using this tool, the authors reviewed and edited the content as needed and take full responsibility for the content of the publication.

\bibliography{ifacconf}             
                                            
\end{document}

%% file: math_commands.tex
\def\de{\mathrm{d}}

\def\ddt{\frac{\de}{\de t}}

%% file: pace_shorthands.tex
\def\udq{\bm{u}_\mathrm{dq}}

\def\uabc{\bm{u}_\mathrm{abc}}

\def\idq{\bm{i}_\mathrm{dq}}
\def\id{{i}_\mathrm{d}}
\def\iq{{i}_\mathrm{q}}
\def\iabc{\bm{i}_\mathrm{abc}}

\def\psidq{\bm{\psi}_\mathrm{dq}}
\def\psid{{\psi}_\mathrm{d}}
\def\psiq{{\psi}_\mathrm{q}}